\documentclass[eqsecnum,twocolumn,showpacs,preprintnumbers,amssymb,aps]{revtex4}

\usepackage{graphicx}
\usepackage{dcolumn}
\usepackage{amsthm,amsmath}
\usepackage{longtable}

\begin{document}
%\preprint{\today}

\title{Implementation and Application of the Relativistic Equation of Motion Coupled-cluster Method for the Excited States of Closed-shell Atomic Systems}
\vspace{0.5cm}

\author{D. K. Nandy \footnote{Email: dillip@prl.res.in}, Yashpal Singh and B. K. Sahoo \footnote{Email: bijaya@prl.res.in}}
\affiliation{Theoretical Physics Division, Physical Research Laboratory, Navrangpura, Ahmedabad 380009, India}

\date{Received date; Accepted date}

\vskip1.0cm

\begin{abstract}

We report the implementation of equation-of-motion coupled-cluster (EOMCC) method in
the four-component relativistic framework with the spherical atomic potential to generate
the excited states from a closed-shell atomic configuration. This theoretical development
will be very useful to carry out high precision calculations of varieties of atomic properties
in many atomic systems. We employ this method to calculate excitation energies of many
low-lying states in a few Ne-like highly charged ions, such as
Cr XV, Fe XVII, Co XVIII and Ni XIX ions, and compare them against their corresponding experimental 
values to demonstrate the accomplishment of the EOMCC implementation. The considered ions are
apt to substantiate accurate inclusion of the relativistic effects in the evaluation of
the atomic properties and are also interesting for the astrophysical studies. Investigation
of the temporal variation of the fine structure constant ($\alpha$) from the astrophysical 
observations is one of the modern research problems for which we also estimate the $\alpha$
sensitivity coefficients in the above ions.

\end{abstract} 

\maketitle

\section{Introduction}

With the advent of sophisticated advanced technologies, modern research in atomic physics demands
many high precision atomic calculations. Some of the prominent examples in this context are, studies of parity 
non-conservation (PNC) and permanent electric dipole moments (EDMs) \cite{ginges,ritz},
estimation of the uncertainties for the frequency standard measurements \cite{hall,hansch,wineland},
probing variation of the fine structure constant \cite{uzan,chand},
extracting nuclear charge radii and nuclear moments \cite{martensson, wansbeek, avgoulea},
providing atomic data for the astrophysical investigations \cite{ivanchik, king} et cetera. In the last
two decades, the coupled-cluster (CC) method for the single valence systems in the
four-component relativistic framework have been extensively employed for the above mentioned
research problems with a great success \cite{bijaya1,bijaya2,knuc,bbr}. In contrast, such CC methods are scanty for calculating the
excited state properties of the systems with closed-shell configurations as far as the 
four-component relativistic approach with the explicit form of the spherical atomic potentials 
are concerned. There have been development of CC methods in the Fock-space formalism to calculate
these states \cite{kaldor1,kaldor2,angom1}, however such approaches suffer from two serious problems: (i) It increases
the computational complexity when applied to estimate the matrix elements of an operator and (ii)
It yields intruder state problem while increasing the size of the model space \cite{bartlett}. On the other hand, 
the low-lying odd parity forbidden transitions among the singlet states of the systems like
Mg, Ca, Sr, Al$^+$, In$^+$, Hg$^+$, Yb etc. are considered for the atomic clock experiments
\cite{diddams,gill1,gill2,rosenband1,rosenband2}. 
Similarly, atomic systems like Xe, Ba, Ra, Yb, Hg, Rn, etc. have been considered 
for the PNC and EDM studies \cite{wood,dzuba,griffith,rosenberry,lathalett,yashpal} for which high precision calculations in the relativistic method
are indispensable. Coincidently, the excited states involved in the above research problems
can be created by exciting one electron from an occupied orbital (hole (h)) to a unoccupied
orbital (particle (p)) which is customarily referred to as 1h-1p excitation in the literature.

 Among various popular many-body methods, equation-of-motion coupled-cluster (EOMCC)
theory is one of the better suited methods to obtain the excited states \cite{bartlett, piecuch}. This method is 
formulated in the similar way as the excitation operators defined for the configuration-interaction (CI) 
method, however the excitations are carried out with respect to the exact state in contrast to the 
model reference state of the CI method. Uniqueness of this approach is that the energy differences 
between the atomic states are estimated directly by casting the Schr\"odinger equations in a particular 
form. Many non-relativistic calculations on the ionization potentials (IPs), electron affinities (EAs)
and excitation energies (EEs) in different atomic and molecular systems have been reported
in the EOMCC framework \cite{bartlett, piecuch, bartlett1, gour, mukherjee}. Recently, this method has been developed 
to determine the first and second IPs of many
closed-shell systems using the atomic integrals in the four-component relativistic mechanics \cite{himadri1, himadri2}.
However, all the above EOMCC calculations are carried out using the molecular codes considering
special group symmetry properties. In this work, we discuss about the implementation of the 
EE determining EOMCC method (referred to as EE-EOMCC method)
based on the four-component relativistic mechanics and expressing the atomic potentials
explicitly in the spherical polar coordinates. Also, all the physical operators are represented
in terms of the Racah angular momentum operators for which we make use of the reduced matrix elements
in order to reduce the computational scalability.

  In order to corroborate the successful implementation of the excited states determining 
EE-EOMCC method, we calculate EEs of transitions among the low-lying states in few
Ne-like systems that are of immense astrophysical interest. Transition lines in the range
10.5-21.2 $\AA$ of Fe XVII are observed in the solar corona \cite{Edlen,Tyren,Klapisch,McKenzie}.
It is also found from the stellar binary Capella observation that Fe XVII is one of the 
important constituents in the stellar corona \cite{Audard} and transition lines from this ion
have also been observed from various astrophysical objects by many others \cite{Dupree, Lepson, 
Sako, Jupen, Dere, Feldman1}. The above lines and transitions from Co XVIII and Ni XIX are also
very important for the astrophysical and tokamak plasma studies \cite{Klapisch,Feldman2}. The
bottom line for the referral of these astrophysical observation is that the above transition
lines could be of potential candidates for probing temporal variation of the fine structure 
constant ($\alpha$). In our recent works, we have provided data to probe $\alpha$ variation by employing
a single reference CC method formulated in the Fock-space approach for the F- and Cl-like Cr, 
Fe and Ni ions \cite{nandy1,nandy2}. We intend, here, to estimate the sensitivity coefficients 
in the considered ions for the investigation of $\alpha$ variation whose magnitudes would
magnitudes would gauge the significance of the relativistic effects and can be undertaken
for their detection.

\section{Perception of $\alpha$ variation sensitivity coefficients} 

Probing different absorption systems, associated with the quasi-stellar objects like quasars, 
can provide useful information regarding the speculated temporal variability of $\alpha$. These
absorption systems that are present at different redshifts contain many metallic and non-metallic
ions of various elements. A small fraction of the absorption systems, that are detected
through the analysis of the
quasars spectra, could be intrinsically associated to the quasars themselves. Investigations of the
broad absorptions line systems (BALs) of this region reveal that the chemical compositions are usually 
from the highly ionized species \cite{behar, green}. Next to BALs, the most important regions are
the intervening absorption systems which can be further classified into various regions depending 
upon the column density of the neutral hydrogen (H) atoms \cite{prochaska, wolfe, sargent}. The 
absorptions lines that are coming out of these systems are red-shifted  due to the cosmological 
expansion of the universe and are related to the cosmological redshift parameter ($z$) by the 
following relation
\begin{eqnarray}
\lambda_z=\lambda_{rest}(1+z),
\end{eqnarray}
where $\lambda_{rest}$ is the wavelength at the time of emission in the rest frame of the absorption system.
With the precise knowledge of the redshift of an absorption line, one can extract the information
about the subtle temporal variation in $\alpha$ after the cautious consideration of the systematic uncertainties
associated with the observation \cite{molaro}. These absorption lines are observed with the advanced telescope
such as high-resolution Echelle Spectrograph (HIRES) at the Keck Observatory or the UV-Visual Echelle Spectrograph
(UVES) at the ESO Very Large Telescope (VLT) for the investigation of variation in $\alpha$ \cite{levshakov1, murphy}. 

The anticipated tiny variation in $\alpha$ from the present laboratory value $\alpha_0$
can be inferred by combining the calculated relativistic sensitivity coefficients ($q$s) 
of different atomic transitions with the observed spectral lines from the quasars \cite{chand}. 
Since the energy of an atomic level scales at the order of  $\alpha^2$ in the relativistic theory, the 
frequency of an atomic transition will depend on the value of $\alpha$ at a given time.
The relativistic corrections to the energy levels of a multi-electron atom can be expressed as \cite{Dzuba1}
\begin{eqnarray}
\Delta=-\frac{Z^2_a}{2}\frac{(Z\alpha)^2}{\nu^3} \left ( \frac{1}{J+1/2}-\frac{Z_a}{Z\nu} \left [ 1-\frac{Z_a}{4Z} \right ] \right ),
\end{eqnarray}
with $Z$ is the atomic number, $J$ is the angular momentum of the state and $\nu$ and $Z_a$ 
are the effective principal quantum number and effective atomic number, respectively, of an
outer electron due to the screening effects of the inner core electrons.
It has been shown that instead of considering two transitions from a particular atomic
system (alkali doublet (AD) method), it is advantageous to compare as many as transitions
from a number of systems (many-multiplet (MM) method) to yield an order of magnitude improvement
in the detection of change in $\alpha$ value ($\Delta \alpha$) from the observations \cite{Dzuba2, Murphy1, Sergei}.
Generally, one compares the measured velocity profile in the MM method to infer tiny shifts 
in the transitions, having different magnitudes of $q$ parameters, to obtain stringent value
of $\frac{\Delta \alpha}{\alpha}$ from a best possible fit.
In this method the change in the transition frequency between two states of an atomic system with respect to an 
arbitrary variation in $\alpha$, quantified as $x=(\frac{\alpha}{\alpha_0})^2-1$, can be 
given by
\begin{eqnarray}
\omega( \alpha^2)\approx\omega(\alpha_0^2)+qx,
\end{eqnarray}
such that $q=\frac{d \omega}{d ( \alpha^2) }|_{x=0}$ corresponds to rate of change of
$\omega$, which is independent of $x$, and known as the sensitivity 
coefficient for $\alpha$ variation. In the MM method, the commonly used relation for the 
extraction of change in $\alpha$ is given by
\begin{eqnarray}
\frac{\Delta v}{c}=-\frac{2q}{\omega(\alpha_0^2)}\frac{\Delta \alpha}{\alpha},
\end{eqnarray}
where $c$ is the velocity of light, $\Delta v$ corresponds to the change in the velocity 
profile of the absorption lines that are related to the wavelengths of the atomic transitions. 
Therefore for probing $\alpha$ variation using the MM method, it is imperative to find out 
$q$ parameters in many possible transitions of the atomic systems that are highly abundant in the 
astrophysical objects like the considered ions in the present work.

\section{Relativistic atomic integrals}\label{sec2}

For the present calculation, we consider following relativistic Dirac-Coulomb (DC) Hamiltonian which is re-scaled
with respect to the rest mass energy of the electrons
\begin{eqnarray}
\nonumber H &=& \sum_i \left [ c\mbox{\boldmath$\alpha$}_i\cdot \textbf{p}_i+(\beta_i -1)c^2 + V_{nuc}(r_i) +
\sum_{j>i} \frac{1}{r_{ij}} \right ] \\
\end{eqnarray}
where $\mbox{\boldmath$\alpha$}_i$ and $\beta_i$ are the usual Dirac matrices, $V_{nuc}(r_i)$ is the nuclear
potential and $\frac{1}{r_{ij}}=\frac{1}{|\textbf r_i - \textbf r_j|}$ is the inter-electronic 
Coulombic repulsion potential. The nuclear potential is evaluated by considering the Fermi-charge distribution of the nuclear density as given by
\begin{equation}
\rho_{nuc}(r)=\frac{\rho_{0}}{1+e^{(r-b)/d}}
\end{equation}
where the parameter `$b$' is the half-charge radius as $\rho_{nuc}(r)=\rho_0/2$ for $r=b$ and `$d$' is related
to the skin thickness which are evaluated by
\begin{eqnarray}
d&=& 2.3/4(ln3) \\
\text{and} \ \ \ \ \ b&=& \sqrt{\frac {5}{3} r_{rms}^2 - \frac {7}{3} d^2 \pi^2}
\end{eqnarray}
with $r_{rms}$ is the root mean square radius of the nucleus.

 In the relativistic quantum mechanics, the four-component Dirac wave function for a single
electron is expressed by
\begin{eqnarray}
 |\phi(r) \rangle = \frac {1}{r} \left ( \begin{matrix} P(r) & \chi_{\kappa,m}(\theta,\phi)  \cr
                                                  iQ(r) & \chi_{-\kappa,m}(\theta,\phi) \cr
                                                        \end{matrix}  \right )
\end{eqnarray}
where $P(r)$ and $Q(r)$ are the large and small components of the wave function respectively. The angular components 
have the following form
\begin{eqnarray}
\chi_{\kappa,m}(\theta, \phi) = \sum_{\sigma=\pm \frac {1}{2}} C( l \sigma j; m-\sigma, \sigma) Y_l^{m-\sigma}(\theta, \phi) \phi_{\sigma}
\end{eqnarray}
where $C(l \sigma j; m-\sigma, \sigma )$ is Clebsch-Gordan (Racah) coefficient, $Y_l^{m-\sigma}(\theta, \phi)$ 
represents normalized spherical harmonics, $\phi_{\sigma}$ serves as the Pauli two-component spinors and the relativistic quantum number
$\kappa = -(j+\frac {1}{2})a$ embodies the total and orbital quantum numbers $j$ and $l = j - \frac{a}{2}$.

With the defined Dirac-Fock (DF) potential as
\begin{eqnarray}
U|\phi_j\rangle = \sum_{a=1}^{occ} \langle \phi_a |\frac{1}{r_{ja}}|\phi_a \rangle |\phi_j \rangle
 - \langle \phi_a |\frac{1}{r_{aj}}|\phi_j \rangle |\phi_a \rangle,
\end{eqnarray}
summed over all the occupied orbitals $occ$,
the DF wave function ($|\Phi_0 \rangle$) for a close-shell atomic system is obtained by solving the
equation 
\begin{eqnarray}
H_{DF} |\Phi_0 \rangle = E_{DF}^{(0)} |\Phi_0 \rangle,
\end{eqnarray}
which in terms of the single particle orbitals are given by
\begin{eqnarray}
\sum_i [ h_0 |\phi(r_i) \rangle = \epsilon_i |\phi(r_i) \rangle ]
\end{eqnarray}
for the DF Hamiltonian
\begin{eqnarray}
H_{DF} &=& \sum_i \left [ c\mbox{\boldmath$\alpha$}_i\cdot \textbf{p}_i+(\beta_i -1)c^2 + V_{nuc}(r_i) +U(r_i) \right ] \nonumber \\
&=& \sum_i h_0(r_i),
\end{eqnarray}
where $h_0$ is the single particle Fock operator.

We express $|\phi_{n,\kappa}(r)\rangle$, with the principal quantum number 
$n$ and angular quantum number $\kappa$, of an electron orbital as linear combination of Gaussian type of 
orbitals (GTOs) to obtain the DF orbitals. In the spherical polar coordinates, it is given by
\begin{eqnarray}
 |\phi_{n,\kappa}(r) \rangle = \frac {1}{r} \sum_{\nu} \left (
         \begin{matrix}
         C_{n,\kappa}^L N_L f_{\nu}(r) & \chi_{\kappa,m} \cr
         i C_{n, -\kappa}^S N_S \left (\frac{d}{dr} + \frac{\kappa}{r} \right ) f_{\nu}(r) &\chi_{-\kappa,m}\cr
                 \end{matrix}
         \right ), \ \
\end{eqnarray}
where $C_{n,\kappa}$s are the expansion coefficients, $N_{L(S)}$ are the normalization constants for the large
(small) components of the wave function and $f_{\nu}(r) = r^l e^{-\eta_{\nu} r^2}$ are the GTOs
with the suitably chosen parameters $\eta_{\nu}$ for orbitals of different angular momentum symmetries. For the exponents, we use the even
tempering condition $\eta_{\nu} = \eta_0 \zeta^{\nu-1}$ with two parameters $\eta_0$ and $\zeta$. It  can be
noticed in the above expression that the large and small components of the wave function satisfy the kinetic balance
condition. The orbitals are finally obtained by executing a self-consistent procedure to solve the following eigenvalue
form of the DF equation 
\begin{eqnarray}
\sum_{\nu} \langle f_{i,\mu}|h_0|f_{i,\nu}\rangle c_{i\nu} 
= \epsilon_i \sum_{\nu} \langle f_{i,\mu}|f_{i,\nu}\rangle c_{i\nu}, 
\end{eqnarray}
which is in the matrix form given by
\begin{eqnarray}
\sum_{\nu} F_{\mu \nu} c_{i \nu} = \epsilon_i \sum_{\mu \nu} \langle f_{i,\mu}|f_{i,\nu}\rangle c_{i \nu}.
\end{eqnarray}
The above equation implies that the parity and the total angular momentum of an orbital are fixed which are the
essential conditions to describe the mechanics in the spherical coordinates.

 To retain the atomic spherical symmetry property in our calculations, the matrix form of the Coulomb
interaction operator using the above single particle wave functions are expressed as
\begin{eqnarray}
\langle \phi_a \phi_b  | \frac {1}{r_{12}} | \phi_c \phi_d \rangle &=& \int dr_1 [P_a(r_1)P_c(r_1) + Q_a(r_1)Q_c(r_1)]
 \nonumber \\ & \times & \int dr_2 [P_b(r_2)P_d(r_2) + Q_b(r_2)Q_d(r_2)] \nonumber \\
  & \times & \sum_k \frac {r_<^k}{r_>^{k+1}}  \times Ang,
\label{ceq}
\end{eqnarray}
in which the $k$ multi-poles are determined by considering the triangle conditions $|j_a - j_c| \le k \le j_a + j_c$ and $|j_b - j_d| \le k \le j_b + j_d$ along with the additional restrictions over $k$ by multiplying a factor $\Pi(\kappa,\kappa',k) = \frac {1}{2} [1-aa'(-1)^{j+j'+k}]$ that is finite only
for $l + l' + k =$ even. The
angular momentum factor of the above expression is given by
\begin{eqnarray}
Ang &=& \delta(m_a-m_c,m_d-m_d) \Pi(\kappa_a,\kappa_c,k) \Pi(\kappa_b,\kappa_d,k) \nonumber \\
  && \times (-1)^q \sqrt{(2j_a+1)(2j_b+1)(2j_c+1)(2j_d+1)} \nonumber \\ &&
   \times \left (
         \begin{matrix} 
         j_a & k & j_c \cr
         -m_a & q & m_c \cr
       \end{matrix}
         \right ) 
   \left (
         \begin{matrix} 
         j_b & k & j_d \cr
         -m_b & -q & m_d \cr
       \end{matrix}
         \right ) \nonumber \\ &&
  \times  \left (
         \begin{matrix} 
         j_a & k & j_c \cr
         \frac {1}{2} & 0 & -\frac {1}{2} \cr
       \end{matrix}
         \right )
   \left (
         \begin{matrix} 
         j_b & k & j_d \cr
         \frac {1}{2} & 0 & -\frac {1}{2}\cr
       \end{matrix}
         \right ), 
\end{eqnarray}
where $m_j$ is the azimuthal component of $j$.
In order to minimize the computational efforts, we use the reduced matrix elements. 
Thus, we express
\begin{eqnarray}
\langle \phi_a \phi_b  | \frac {1}{r_{12}} | \phi_c \phi_d \rangle &=& 
 \delta(m_a-m_c,m_d-m_b) \sum_{k,q} \Pi(\kappa_a,\kappa_c,k) \nonumber \\ & \times & \Pi(\kappa_b,\kappa_d,k) 
(-1)^{j_a-m_a+j_b-m_b+k-q}  \nonumber \\ & \times & \left (
         \begin{matrix} 
         j_a & k & j_c \cr
         -m_a & q & m_c \cr
       \end{matrix}
         \right )
   \left (
         \begin{matrix} 
         j_b & k & j_d \cr
         -m_b & -q & m_d \cr
       \end{matrix}
         \right ) \nonumber \\ & \times & \langle ab || \frac{1}{r_{ij}} || cd \rangle, 
\end{eqnarray}
with the reduced matrix element
\begin{eqnarray}
 \langle ab || \frac{1}{r_{ij}} || cd \rangle &=& (-1)^{j_a+j_b+k+1} \nonumber \\ & \times & \int dr_1 [P_a(r_1)P_c(r_1) + Q_a(r_1)Q_c(r_1)]
 \nonumber \\ & \times & \int dr_2 [P_b(r_2)P_d(r_2) + Q_b(r_2)Q_d(r_2)]  \nonumber \\ & \times & 
  \sqrt{(2j_a+1)(2j_b+1)(2j_c+1)(2j_d+1)} \nonumber \\ & \times &
   \frac {r_<^k}{r_>^{k+1}}  
  \left (
         \begin{matrix} 
         j_a & k & j_c \cr
         \frac {1}{2} & 0 & -\frac {1}{2} \cr
       \end{matrix}
         \right )
   \left (
         \begin{matrix} 
         j_b & k & j_d \cr
         \frac {1}{2} & 0 & -\frac {1}{2}\cr
       \end{matrix}
         \right ).
\end{eqnarray}

\section{Relativistic EE-EOMCC method for atoms}

\begin{center}
\begin{figure}[t]
\includegraphics[width=7.0cm, height=2.2cm, clip=true]{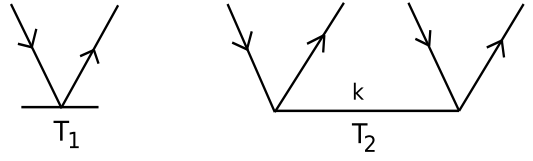}
\caption{Diamagnetic representations of the $T_1$ and $T_2$ excitation operators for the closed-shell CC method. $k$ is 
the rank of the $T_2$ operator.} 
\label{fig1}
\end{figure}
\end{center}

 The starting point of our EOMCC method is the ground state wave function ($|\Psi_0\rangle$) of a closed-shell
system which in the CC formalism is expressed as 
\begin{eqnarray}
|\Psi_0\rangle= e^T |\Phi_0 \rangle,
\end{eqnarray}
where $|\Psi_0\rangle$ is the exact ground state and $|\Phi_0 \rangle$ is the DF reference state taken in the
anti-symmetrized form. We have restricted to only the singly and doubly excited configurations from $|\Phi_0 \rangle$ 
in our calculations (CCSD method) by defining $T=T_1 + T_2$, which in the second quantization notation are given by
\begin{eqnarray}
   T_1 =\sum_{a,p}a^{\dagger}_p a_a t^p_a, 
\ \ \ \text{and} \ \ \ T_2 =\frac{1}{4}\sum_{ab,pq}a^{\dagger}_pa^{\dagger}_qa_ba_a t^{pq}_{ab},
\end{eqnarray}
where the subscripts $a,b$ and $p,q$ represent for the core and virtual orbitals, $a$ and $a^{\dagger}$ are the annihilation and
creation operators, and $t_a^p$ and $t_{ab}^{pq}$ are the singly and doubly excited amplitudes. In a spherical 
coordinate system they are expressed as
\begin{eqnarray}
\nonumber \langle jm_j|T_1|j'm_j^{'} \rangle 
                                 &=&  (-1)^{j-m_j} \sum_{k,q} \left (
                                           \begin{matrix}
                                           j & k & j' \cr
                                           -m_j & q & m_j^{'} \cr
                                           \end{matrix}
                                         \right )
                                         \langle j||t_1^k||j' \rangle \\
\end{eqnarray}
and

\begin{eqnarray}
&& \langle j_a m_a; j_b m_b |T_2| j_c m_c; j_d m_d \rangle = (-1)^{j_a -m_a + j_b -m_b}  \nonumber \\ \nonumber
&&\times \sum_{k,q} (-1)^{k-q}  \times    \left (
         \begin{matrix}
         j_a & k & j_c \cr
         -m_a & q & m_c \cr
         \end{matrix}
         \right )  
   \left (
         \begin{matrix}
         j_b & k & j_d \cr
         -m_b & -q & m_d \cr
         \end{matrix}
         \right ) \\ \nonumber
&& \times        \langle j_a j_b ||t_2^k||j_c j_d \rangle,
\end{eqnarray}
where $\langle j||t_1^k||j' \rangle$ and $\langle j_a j_b ||t_2^k||j_c j_d \rangle$ are the reduced 
matrix elements of the $T_1$ and $T_2$ operators, respectively. Owing to the nature of our orbitals,
the $T_1$ operator is scalar in our calculations but $T_2$ will have multi-poles
satisfying the triangle conditions $|j_a - j_c| \le k \le j_a+j_c$ and $|j_b-j_d| \le k\le j_b+j_d$. 
Following Eq. (\ref{ceq}), it is evident that the multi-poles satisfying the conditions
$l_a + l_c + k =$ even and $l_b + l_d + k =$ even will be the dominant contributing multi-poles.
Diagrammatic representations of the $T_1$ and $T_2$ operators are shown in Fig. \ref{fig1}.

The above singles and doubles CC amplitude equations are solved using the following matrix form
\begin{eqnarray}
\langle \Phi_0^* || H_N^{eff} \otimes T^* ||\Phi_0 \rangle &=& 0,
\label{eqn1}
\end{eqnarray}
where the superscript $|\Phi_0^* \rangle$ corresponds to the singles ($|\Phi_1 \rangle$) and doubles ($|\Phi_2 \rangle$) excited configurations
from $|\Phi_0 \rangle$ and $H_N^{eff} = (H_N e^T)^{op}_c$ is the effective normal ordered Hamiltonian 
containing only the connected ($c$) open (op) terms.  Here, $T^*$s are the $T_1$ and $T_2$ operators
in the singles and doubles amplitude solving equations respectively.

\begin{center}
\begin{figure}[t]
\includegraphics[width=7.0cm, height=2.4cm, clip=true]{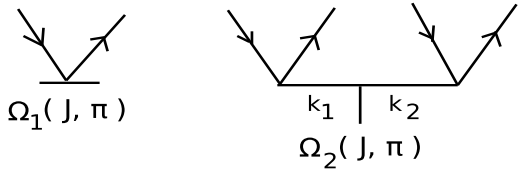}
\caption{Diamagnetic representations of the $\Omega_1$ and $\Omega_2$ EE-EOMCC excitation operators.
 $J$ and $\pi$ are the total angular momentum and parity carried out by the operators. As shown, $J$ value of the 
$\Omega_2$ operator is determined following a triangular condition among two other operators having ranks $k_1$ and $k_2$.} 
\label{fig2}
\end{figure}
\end{center}
 
 The excited states ($|\Psi_K(J, \pi) \rangle$) having specific total angular momentum $J$ and parity $\pi$ ($=(-1)^l$ with the $l$ orbital quantum number)
from the ground state ($|\Psi_0 \rangle$) of a closed-shell atomic system in an EOMCC method is determined
by defining an excitation operator $\Omega_K$ as
\begin{eqnarray}
 |\Psi_K(J, \pi) \rangle = \Omega_K(J, \pi) |\Psi_0\rangle ,
\end{eqnarray}
where $K=0$, $K=1$, $K=2$ etc. correspond to the ground, singly, doubly excited etc. states respectively.
Analogous to the CC excitation $T$ operators, we express the $\Omega_K$ operators in the second quantized 
notations as
\begin{eqnarray}
\Omega_K &=& \Omega_1 + \Omega_2 + \cdots \nonumber \\
  &=& \sum_{a,p} \omega_a^p a^{p^{\dagger}} a_a + \frac {1}{4} \sum_{ab,pq}
  \omega_{ab}^{pq} a^{p^{\dagger}} a^{q^{\dagger}} a_b a_a + \cdots ,
\end{eqnarray}
where $\omega_a^p$, $\omega_{ab}^{pq}$, etc. are the amplitudes of the $\Omega_1$, $\Omega_2$ etc.
operators and obviously, here, $\Omega_0=1$. Thus, the eigenvalue equation for the excited states 
are given by
\begin{eqnarray}
H |\Psi_K \rangle = E_K |\Psi_K\rangle \nonumber \\
H \Omega_K e^T |\Phi_0 \rangle = E_K \Omega_K e^T |\Phi_0\rangle .
\end{eqnarray}

\begin{center}
\begin{figure}[t]
\includegraphics[width=8.0cm, height=5.0cm, clip=true]{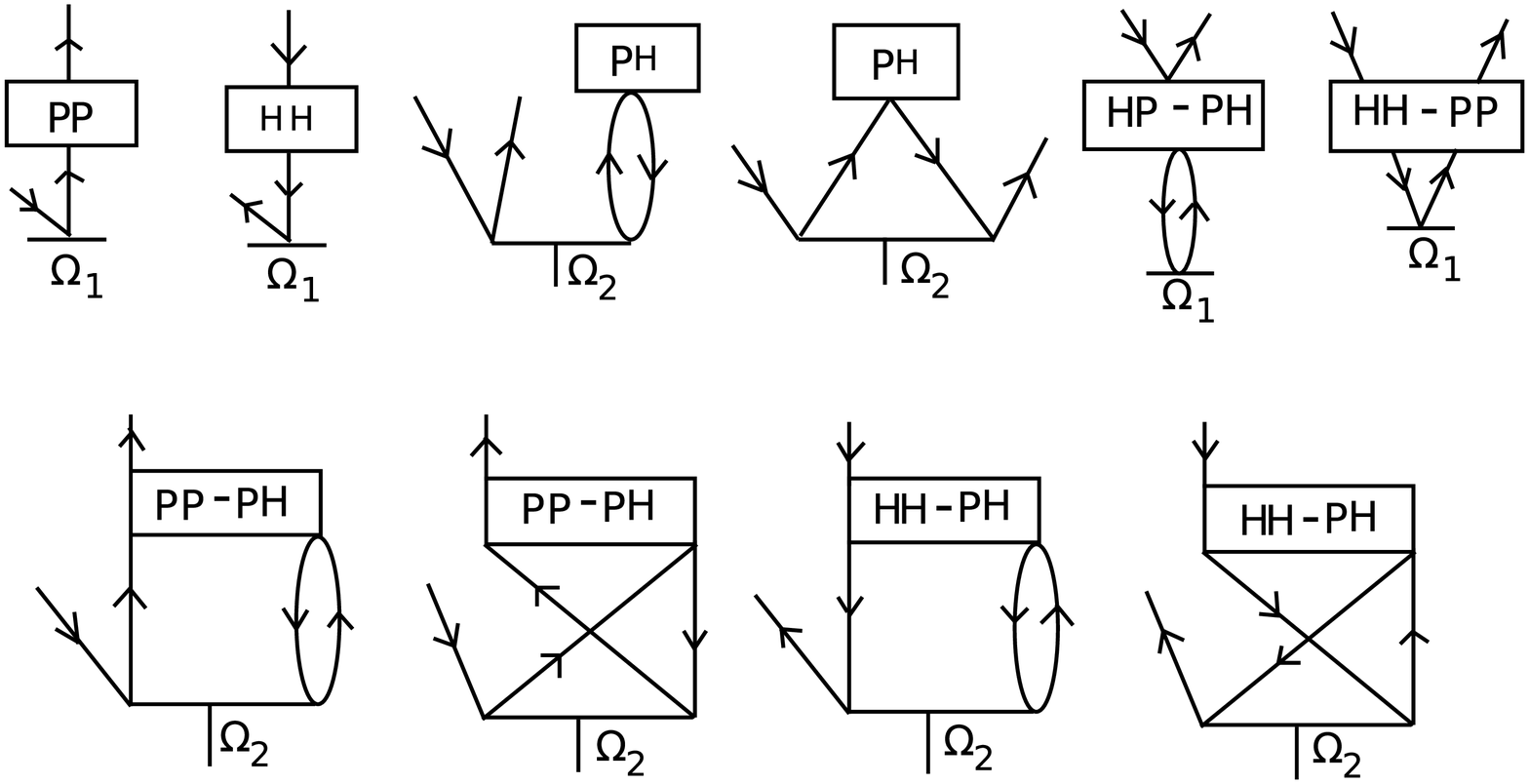}
\caption{EE-EOMCC diagrams to determine amplitudes of the $\Omega_1$ operators.} 
\label{fig3}
\end{figure}
\end{center}
 
 Following the second quantization notations, we can show that $\Omega_K$ and $T$ commute each other. 
Therefore by operating $e^{-T}$ from the left side of the above equation, we get
\begin{eqnarray}
e^{-T} H e^T \Omega_K |\Phi_0 \rangle &=& E_K \Omega_K |\Phi_0 \rangle
\nonumber \\
(e^{-T} H_N e^T +E_{DF}) \Omega_K |\Phi_0 \rangle &=& E_K \Omega_K |\Phi_0
\rangle \nonumber \\
\{(H_N e^T)_c^{op} +E_g\} \Omega_K |\Phi_0 \rangle &=& E_K
\Omega_K |\Phi_0 \rangle \nonumber \\
H_N ^{eff} \Omega_K |\Phi_0 \rangle &=& \Delta E_K \Omega_K
|\Phi_0 \rangle ,
\label{eq4.9}
\end{eqnarray}
where $E_{DF}(=\langle \Phi_0 | H|\Phi_0\rangle)$ and $E_g$ are the DF
and ground state energies, respectively. Therefore, $\Delta E_K(=E_K-E_g)$  corresponds to the
excitation energy of the $|\Psi_K\rangle$ state with respect to the ground state. 
Using the effective Hamiltonian $H_N^{eff}$, we evaluate the excitation energies 
after projecting $\langle \Phi_L |$ from the left hand side to yield in the following form
\begin{eqnarray}
\langle \Phi_L | H_N^{eff} \Omega_K |\Phi_0 \rangle = \Delta
E_K \langle \Phi_L | \Omega_K |\Phi_0 \rangle \delta_{L,K},
\end{eqnarray}
where $|\Phi_L\rangle$ represents an excited determinantal state with definite values of $J$ and $\pi$.
Therefore, $\Omega_K$s have the fixed $J$ and $\pi$ values for 
which we get
\begin{widetext}
\begin{eqnarray}
\langle \Phi_L (J, \pi) | H_N^{eff}  \Omega_K (J, \pi) |\Phi_0 \rangle &=& \Delta E_L \langle
\Phi_L (J, \pi) | \Omega_L (J, \pi) |\Phi_0 \rangle .
\end{eqnarray}

By applying the completeness principle, the above equation corresponds to
\begin{eqnarray}
\sum_K \langle \Phi_L (J, \pi) | H_N^{eff} | \Phi_K \rangle \langle \Phi_K | \Omega_K (J, \pi) |\Phi_0 \rangle 
= \Delta E_L \langle \Phi_L (J, \pi) | \Omega_L (J, \pi) |\Phi_0 \rangle
\end{eqnarray}

Considering only the singles ($\Omega_1$) and doubles ($\Omega_2$) excitations only, we write down the matrix form as
\begin{eqnarray}
\left ( \begin{matrix}  \langle \Phi_1 (J, \pi) | H_N^{eff} | \Phi_1 (J, \pi) \rangle & \langle
\Phi_1 (J, \pi) | H_N^{eff} | \Phi_2 (J, \pi) \rangle \cr
\langle \Phi_2 (J, \pi) | H_N^{eff} | \Phi_1 (J, \pi) \rangle & \langle \Phi_2 (J, \pi) | H_N^{eff} |
\Phi_2 (J, \pi) \rangle \cr
\end{matrix} \right ) \left ( \begin{matrix}  \langle \Phi_1 (J, \pi) | \Omega_1 (J, \pi) | \Phi_0 \rangle \cr
\langle \Phi_2 (J, \pi) | \Omega_2 (J, \pi) | \Phi_0 \rangle \end{matrix} \right ) &=& 
\Delta E_{1/2}
\left ( \begin{matrix}  \langle \Phi_1 (J, \pi) | \Omega_1 (J, \pi) | \Phi_0 \rangle \cr
\langle \Phi_2 (J, \pi) | \Omega_2 (J, \pi) | \Phi_0 \rangle \end{matrix} \right ) . \nonumber \\
\end{eqnarray}
\end{widetext}

\begin{center}
\begin{figure}[t]
\includegraphics[width=8.3cm, height=6.8cm, clip=true]{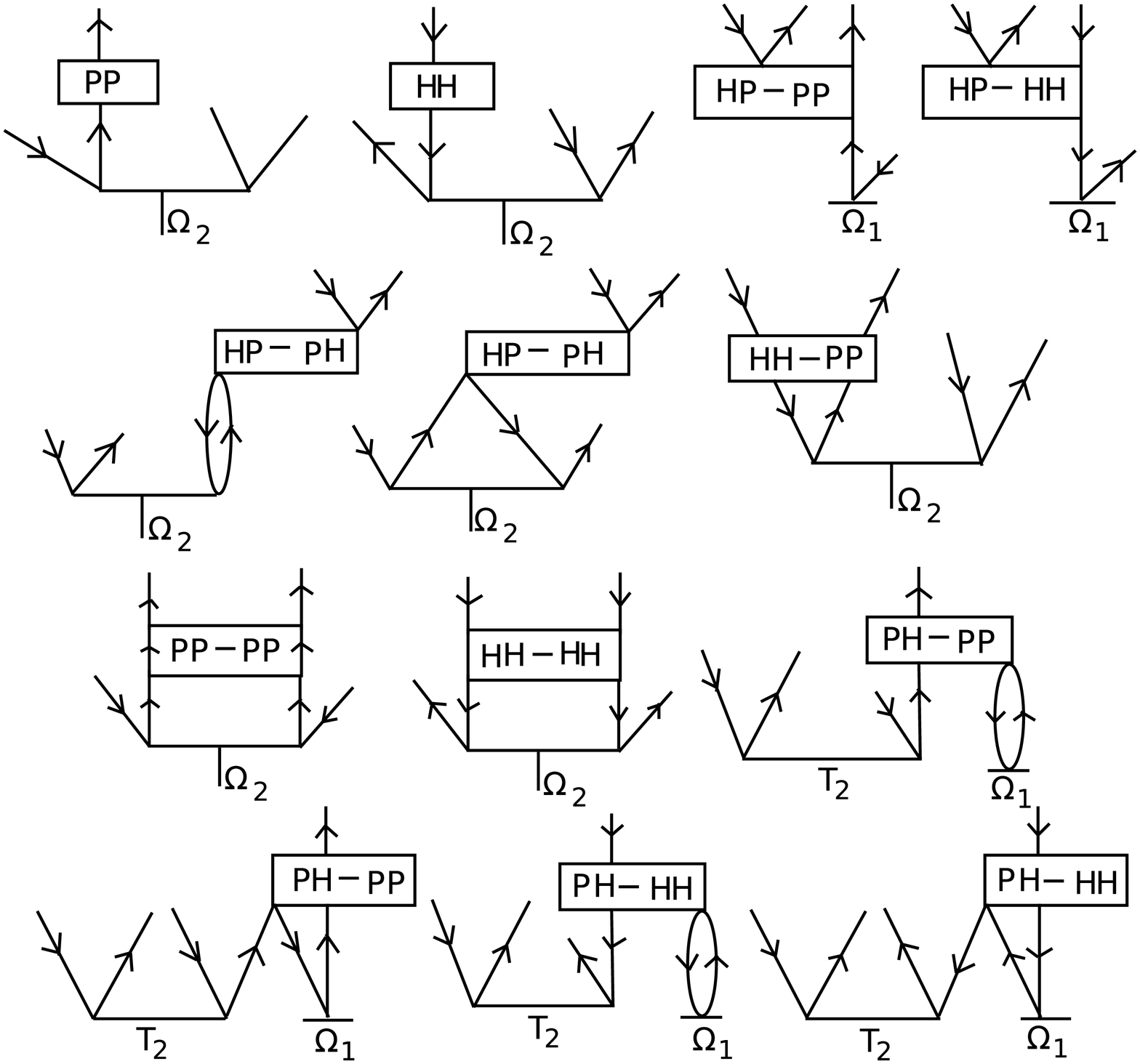}
\caption{EE-EOMCC diagrams to determine amplitudes of the $\Omega_2$ operators.} 
\label{fig4}
\end{figure}
\end{center}
 
 The above matrix is non-symmetric in nature with all the finite matrix elements. We use
a modified Davidson algorithm in a iterative scheme to obtain only few roots of the lower 
eigenvalues as has been applied in \cite{kallay}. Unlike the $T$ operators, $\Omega_1$ has
a finite rank equal to $J$ of the state. Similarly, $\Omega_2$ has the effective rank equal
to $J$ but in contrast to the $T_2$ operator, which is obtained from the scalar product
of two equal ranked tensor operators, the $\Omega_2$ operators are the outcome of general
tensor product between two arbitrary ranked tensors. Therefore, its final rank $J$ has been 
determined using the following types of products
\begin{eqnarray}
\langle J_1 \pi || [t^{k_1} u^{k_2}]^J || J_2 \pi  \rangle = (2J+1)^{1/2} \sum_{J_3} (-1)^{J_1+J_2+J}  \nonumber \\
                                  \left \{ 
                                           \begin{matrix}
                                           k_1 & k_2 & J \cr
                                           J_2 & J_1 & J_3 \cr
                                           \end{matrix}
                                         \right \}
                                         \langle J_1 \pi ||t^{k_1} ||J_3  \pi \rangle \langle J_3  \pi ||u^{k_2} ||J_2  \pi \rangle .
\end{eqnarray}
The diagrammatic representation of the $\Omega_K$ operators are shown in Fig. \ref{fig2}.

To reduce the computational scalability, 
 we divide $H_N^{eff}$ into effective one-body, two-body and three-body terms
with two, four and six open lines, respectively. We also make use of the effective
two-body terms to construct the effective three-body terms. Using these diagrams
the final contributing diagrams to calculate amplitudes for the $\Omega_1$ and
$\Omega_2$ operators are shown in Figs. \ref{fig3} and \ref{fig4}, respectively.

\section{Results and discussion}

 We employ the developed relativistic EE-EOMCC method to calculate energies for many low-lying
excited states of the Cr XV, Fe XVII, Co XVIII and Ni XIX ions with different values of the
total angular momentum and odd parity. The calculated energies at the CCSD level are reported
and compared against the values listed in the national institute of science and technology (NIST) database 
\cite{NIST} in Table \ref{tab1}. In order to realize the role of the correlation effects 
incorporated through the CCSD method, we take the approximation in the effective Hamiltonian 
as $H_N^{eff}\equiv H_N$ and obtain
the EEs in the equation-of-motion framework. These results are quoted as second order
many-body perturbation theory [MBPT(2) method] results in the same table. We also give EEs
estimated from the orbital energies in the same table as the DF results. 

  As seen from Table \ref{tab1}, EEs obtained using the DF method are over estimated from
the experimental results listed in the NIST database and the MBPT(2) results are under
estimated for all the calculated states in the considered ions. Our CCSD results and the 
values from NIST are in close agreement and the differences between them are quoted in terms
of percentage as $\Delta$ in the above table. From the $\Delta$ values, it is clear that the 
percentages of accuracies in our calculations are sub-decimal in all the cases.

Among the theoretical calculations, the most recent one is carried out by Aggarwal et al. 
in which they have employed the multi-configurational Dirac-Fock (MCDF) method using the 
general-purpose relativistic atomic structure package (GRASP) \cite{grasp}
to calculate the energies along with other properties of Fe XVII ion \cite{Aggarwal}.
Their calculated energies are found to be under estimated in all the considered excited states
compared to the NIST data. Bhatia et al. had calculated some of these energies in Fe XVII using 
a SuperStructure (SS) code in a semi-relativistic approach considering the Breit-Pauli 
Hamiltonian \cite{Bhatia}, but their results, when compared with the NIST data, are slightly over 
estimated and less accurate than our CCSD results. Further, Cornille et al. had also evaluated these 
energy levels along with few more states using the same SS code \cite{Cornille} but their estimated 
energies are undervalued than the NIST results. In another work, Sampson et al. had obtained the excitation 
energies in the Fe XVII ion using a Dirac-Fock-Slater (DFS) atomic code \cite{Sampson} and their 
results follow a similar trend as obtained by Bhatia et al.. As compared to others, our CCSD results for 
the EEs in the Fe XVII ion are in close agreement with the values given in the NIST database. However,
we do not find calculations of EEs in other ions using the relativistic methods to make a comparative
study. Nevertheless, the excellent agreements between our CCSD results and the experimental values against
the calculations carried out using other methods demonstrate the potential of the EE-EOMCC method to produce 
accurate results for the excited states in the considered closed-shell ions. 

\begin{table*}[t]
\caption{\label{4p4d}Absolutes values of the excitation energies (in $cm^{-1}$) of few low-lying states in the Cr XV, Fe XVII, Co XVIII and Ni XIX ions. }
\begin{ruledtabular}
\begin{center}
\begin{tabular}{lcccccccccc}
  State  &  Term   & J   & Others & \multicolumn{3}{c}{This work} & &  \\
 \cline{5-7} & & \\
       &         &    &      & DF  &MBPT(2)&CCSD&NIST & $\Delta$ \\
\hline \\
Cr XV &   &  \\                             
$2s^22p^6$&$ \ ^1S_0  $       &0 &    &0 &0          &0            &0         &               \\

$2s^22p^53s$  &$\ ^3P^o_2$    &2 &    &  5076633.27 & 4700197.15  &4711049.38&4714294&0.06  \\
              &$\ ^1P^o_1$    &1 &    &  5076633.27 & 4717618.94  &4728787.50&4727500&0.03  \\
$2s^22p^53s$  &$\ ^3P^o_0$    &0 &    &  5150041.38 & 4773037.02  &4783974.23&4784174&0.004  \\
              &$\ ^3P^o_1$    &1 &    &  5150041.38 & 4786118.37  &4796971.89&4793200&0.07  \\
$2s^22p^53d$  &$\ ^3P^o_0$    &0 &    &  5666530.62 & 5241502.53  &5252079.78&5253448&0.03  \\
              &$\ ^3P^o_1$    &1 &    &  5666530.62 & 5250734.95  &5261208.38&5259419&0.03  \\
$2s^22p^53d$  &$\ ^3P^o_2$    &2 &    &  5666530.62 & 5263315.31  &5273858.56&5270945&0.05  \\
\hline \\

Fe XVII &  & \\
$2s^22p^6$                  &$\ ^1S_0  $       &0         & 0     &0 &0          &0            &0         &                       \\

$2s^22p^53s$                &$\ ^3P^o_2$       &2 &5833877.08$^a$ &  6252544.54 & 5837409.29  &5846872.18&5849490&0.04  \\
                            &                  &  &5852291.01$^b$ &             &             &          &       &       \\
                            &                  &  &5838683.58$^c$ &             &             &          &       &       \\
                            &                  &  &5852261.00$^d$ &             &             &          &       &       \\
                            &$\ ^1P^o_1$       &1 &5849646.34$^a$ &  6252544.54 & 5856420.08  &5866053.91&5864770&0.02  \\
                            &                  &  &5868334.60$^b$ &             &             &          &       &       \\
                            &                  &  &5854891.78$^c$ &             &             &          &       &       \\
                            &                  &  &5868270.00$^d$ &             &             &          &       &       \\
$2s^22p^53s$                &$\ ^3P^o_0$       &0 &5935877.92$^a$ &  6358361.47 & 5942319.60  &5951894.77&5951210&0.01  \\
                            &                  &  &5952931.10$^b$ &             &             &          &       &       \\
                            &                  &  &5943131.56$^c$ &             &             &          &       &       \\
                            &                  &  &5953692.00$^d$ &             &             &          &       &       \\
                            &$\ ^3P^o_1$       &1 &5945710.38$^a$ &  6358361.47 & 5955899.18  &5965332.14&5960870&0.07  \\
                            &                  &  &5963268.35$^b$ &             &             &          &       &       \\
                            &                  &  &5953699.26$^c$ &             &             &          &       &       \\
                            &                  &  &5964194.00$^d$ &             &             &          &       &       \\
$2s^22p^53d$                &$\ ^3P^o_0$       &0 &6448998.63$^a$ &  6930674.16 & 6454603.69  &6463808.10&6463980&0.002  \\
                            &                  &  &6468773.30$^b$ &             &             &          &       &       \\
                            &                  &  &6454156.28$^c$ &             &             &          &       &       \\
                            &                  &  &6466891.00$^d$ &             &             &          &       &       \\
                            &$\ ^3P^o_1$       &1 &6456855.82$^a$ &  6930674.16 & 6465750.34  &6474902.24&6471800&0.05  \\
                            &                  &  &6476685.35$^b$ &             &             &          &       &       \\
                            &                  &  &6462628.00$^c$ &             &             &          &       &       \\
                            &                  &  &6475529.00$^d$ &             &             &          &       &       \\
                            &$\ ^3P^o_2$       &2 &6471845.94$^a$ &  6930674.17 & 6481571.54  &6490752.07&6486400&0.07  \\
                            &                  &  &6491971.76$^b$ &             &             &          &       &       \\
                            &                  &  &6478035.12$^c$ &             &             &          &       &       \\
                            &                  &  &6491737.00$^d$ &             &             &          &       &       \\
\hline   \\
Co XVIII &  & \\ 
$2s^22p^6$                  &$ \ ^1S_0      $    &0 &    & 0           &0            &0         &0             \\
$2s^22p^53s$                &$ \ ^3P^o_1  $      &1 &    &  6885258.00 & 6468217.60  &6482537.35&6477900&0.07  \\
$2s^22p^53s$                &$ \ ^1P^o_1  $      &1 &    &  7010851.48 & 6587125.90  &6601429.30&6592400&0.13  \\
$2s^22p^53d$                &$ \ ^3P^o_1  $      &1 &    &  7608386.67 & 7116091.56  &7130032.89&7122000&0.11  \\
$2s^22p^53d$                &$ \ ^3D^o_1  $      &1 &    &  7612379.83 & 7206635.96  &7219885.94&7210800&0.12  \\
$2s^22p^53d$                &$ \ ^1P^o_1  $      &1 &    &  7733980.15 & 7331833.06  &7343682.69&7334600&0.12  \\
\hline \\

Ni XIX  &  &  \\
$2s^22p^6$                   &$ \ ^1S_0  $  &0 &    &0            &0         &0       &0         &               \\

$2s^22p^53s$                 &$\ ^3P^o_2$   &2 &    &  7547742.28 & 7093993.29  &7102363.09&7105260 &0.04  \\
                             &$\ ^1P^o_1$   &1 &    &  7547742.27 & 7114641.30  &7123258.99&7122600 &0.01  \\
$2s^22p^53s$                 &$\ ^3P^o_0$   &0 &    &  7695794.44 & 7240746.48  &7249098.04&7247700 &0.02  \\
                             &$\ ^3P^o_1$   &1 &    &  7695794.44 & 7254858.78  &7263159.00&7258100 &0.07  \\
$2s^22p^53d$                 &$\ ^3P^o_0$   &0 &    &  8316575.42 & 7789854.59  &7797906.92&7797965 &0.001  \\
                             &$\ ^3P^o_1$   &2 &    &  8316575.42 & 7822402.07  &7830453.36&7847100 &0.20  \\
                             &$\ ^3P^o_2$   &1 &    &  8316575.42 & 7803113.62  &7811208.09&7807700 &0.04  \\

\end{tabular}
\end{center}
\end{ruledtabular}
\label{tab1}
$^a$\cite{Aggarwal}, $^b$\cite{Bhatia}, $^c$\cite{Sampson}, $^d$\cite{Cornille}
\end{table*}

After achieving high precision calculations of the energies for many transitions in the considered 
ions, we intend now to estimate the relativistic sensitivity $q$ coefficients for all these 
transitions. In the Fe XVII ion, we have determined the $q$ parameters for 25 possible inter-combination
transitions that are given in Table \ref{tab2}. Among them three transitions 
$2s^22p^53s\ ^3P^o_2\rightarrow 2s^22p^53s\ ^1P^o_1$, $2s^22p^53d\ ^3P^o_0\rightarrow 2s^22p^53d\ ^3P^o_2 $ 
and $2s^22p^53d\ ^3P^o_1 \rightarrow 2s^22p^53d\ ^3P^o_2$ lie in the optical regime with the wavelengths 6544.50 \AA, 
4460.30 \AA \ and 6849.31 \AA \ respectively. There are also two transitions 
$2s^22p^53s\ ^3P^o_0\rightarrow 2s^22p^53s\ ^3P^o_1$ and 
$2s^22p^53d\ ^3P^o_0\rightarrow 2s^22p^53d\ ^3P^o_1$ that lie near infrared regime 
while the rest of the transitions fall within ultraviolet (UV) to extreme ultraviolet (EUV) region of the electromagnetic 
spectrum. It is worth mentioning that the spectra of Fe XVII ion have been extensively studied by many astrophysics 
groups for investigating different astrophysical plasma, solar plasma, and also in the observation of the
absorption lines coming out from various quasars like IRAS 13349+2438 that are detected by the
XMM-Newton observatory \cite{Sako}. Therefore, the above estimated $q$ parameters will serve as the
useful ingredients if the astrophysical observations in these lines are directed 
towards probing temporal variation of the fine structure constant. For the completeness in the
understanding of the numerical results for the estimation of the $q$ parameters, we also present
the EEs as $\omega(+0.025)$ and $\omega(-0.025)$ in the same table for two different 
values of $x$ as $+0.025$ and $-0.025$ referring to two different values of $\alpha$.
The remarkable findings from these results are that we obtain large $q$ parameters with opposite 
signs in different transitions which is, in fact, a very useful criteria to enhance the effect 
indicating the variation in $\alpha$ from the observations of these atomic spectra.  
By analyzing the results of the $q$-parameters in the Fe XVII ion one can find that there are three 
transitions which could be used as an anchor lines, whose wavelengths are insensitive to the variation 
of $\alpha$, and transitions having large $q$-parameters can be used as probe lines, whose 
wavelengths are highly sensitive to variation of $\alpha$ \cite{kozlov}. The $q$ values for these
possible anchor lines are $-1660.20$, $-758$ and $1364.30$ in $cm^{-1}$, whose corresponding laboratory wavelengths 
lie in the EUV, optical and near infrared (NIR) domain of the electromagnetic spectrum. 
Observations of these EUV lines from any absorption system will be red-shifted towards the optical 
range of the spectrum, which could be easily detected using an earth based observatory.
Similarly, we estimate $q$ parameters for 15 possible transitions in the Co XVIII ion and 
present them in Table \ref{tab2}. Unlike the case of Fe XVII, the considered transition frequencies 
in the Co XVIII ion lie only in the UV range. For Co XVIII we find $5$ transition with positive 
$q$-values and the rest $10$ transitions correspond to negative $q$-values. Among all the transitions in Co XVIII 
we can choose $2s^22p^6 \ ^1S_0 \rightarrow 2s^22p^53d \ ^3P^o_1$ transition as an anchor line because 
of its smaller $q$-value.

\begin{table*}[t]
\caption{Sensitivity $q$ coefficients (in $cm^{-1}$) for the Fe XVII and Co XVIII ions using the CCSD method.
The frequencies $\omega (+0.025)$ and $\omega (+0.025)$ are given as absolute values.}
\begin{ruledtabular}
\begin{tabular}{lccccc}
\multicolumn{1}{c}{Transitions} &
\multicolumn{1}{c}{$J_f$}&
\multicolumn{1}{c}{$\lambda (\AA)$}&
\multicolumn{1}{c}{$\omega(+0.025)$} &
\multicolumn{1}{c}{$\omega(-0.025)$}&
\multicolumn{1}{c}{$q$}\\
\hline\\ 
Fe XVII \\
$2s^22p^6 \ ^1S_0              \rightarrow  2s^22p^53s \ ^3P^o_2$&2 &17.09&5845928.04  &5847815.39 &-37747.00    \\   
$~~~~~~~~~~~~~~                \rightarrow  2s^22p^53s \ ^1P^o_1$&1 &17.05&5865127.95  &5866977.40 &-36989.00    \\
$~~~~~~~~~~~~~~                \rightarrow  2s^22p^53s \ ^3P^o_1$&1 &16.77&5967034.92  &5963673.14 &67235.60    \\   
$~~~~~~~~~~~~~~                \rightarrow  2s^22p^53d \ ^3P^o_1$&1 &15.45&6474858.88  &6474941.91 &-1660.60    \\
$~~~~~~~~~~~~~~                \rightarrow  2s^22p^53d \ ^3P^o_2$&2 &15.41&6490842.94  &6490653.18 &-3795.20    \\

$2s^22p^53s\ ^3P^o_2           \rightarrow 2s^22p^53s \ ^1P^o_1 $&1 &6544.50&19199.91   &19162.01  &-758.00    \\
$~~~~~~~~~~~~~~~~~~            \rightarrow 2s^22p^53s \ ^3P^o_0 $&0 &983.09&107655.55  &102338.19 &-106347.20    \\
$~~~~~~~~~~~~~~~~~~            \rightarrow 2s^22p^53s \ ^3P^o_1 $&1 &897.83&121106.88  &115857.75 &-104982.60    \\   
$~~~~~~~~~~~~~~~~~~            \rightarrow 2s^22p^53d \ ^3P^o_0 $&0 &162.73&617721.12  &616148.30 &-31456.40    \\
$~~~~~~~~~~~~~~~~~~            \rightarrow 2s^22p^53d \ ^3P^o_2 $&1 &160.69&628930.84  &627126.52 &-36086.40    \\
$~~~~~~~~~~~~~~~~~~            \rightarrow 2s^22p^53d \ ^3P^o_2 $&2 &157.00&644914.90  &642837.79 &-41542.20    \\

$2s^22p^53s\ ^1P^o_1           \rightarrow 2s^22p^53s \ ^3P^o_0 $&0 &1156.87&88455.64   &83176.18  &-105589.20    \\
$~~~~~~~~~~~~~~~~~~            \rightarrow 2s^22p^53s \ ^3P^o_1 $&1 &1040.58&101906.97  &96695.74  &-104224.60    \\   
$~~~~~~~~~~~~~~~~~~            \rightarrow 2s^22p^53d \ ^3P^o_0 $&0 &166.73 &598521.21  &596986.29 &-30698.40    \\
$~~~~~~~~~~~~~~~~~~            \rightarrow 2s^22p^53d \ ^3P^o_1 $&1 &164.73 &609730.93  &607964.51 &-35328.40    \\
$~~~~~~~~~~~~~~~~~~            \rightarrow 2s^22p^53d \ ^3P^o_2 $&2 &160.16 &625714.99  &623675.78 &-40784.20    \\

$2s^22p^53s\ ^3P^o_0           \rightarrow 2s^22p^53s \ ^3P^o_1 $&1 &10351.97&13451.33   &13519.56  &1364.60    \\   
$~~~~~~~~~~~~~~~~~~            \rightarrow 2s^22p^53d \ ^3P^o_1 $&1 &192.09&521275.29  &524788.33 &70260.80    \\
$~~~~~~~~~~~~~~~~~~            \rightarrow 2s^22p^53d \ ^3P^o_2 $&2 &186.85&537259.35  &540499.60 &64805.00    \\

$2s^22p^53s\ ^3P^o_1           \rightarrow 2s^22p^53d \ ^3P^o_0 $&0 &198.76&496614.24  &500290.55 &73526.2    \\
$~~~~~~~~~~~~~~~~~~            \rightarrow 2s^22p^53d \ ^3P^o_1 $&1 &195.72&507823.96  &511268.77 &68896.20    \\
$~~~~~~~~~~~~~~~~~~            \rightarrow 2s^22p^53d \ ^3P^o_2 $&2 &190.28&523808.02  &526980.04 &63440.40    \\

$2s^22p^53d\ ^3P^o_0           \rightarrow 2s^22p^53d \ ^3P^o_1 $&1 &12787.72&11209.72   &10978.22  &-4630.00    \\
$~~~~~~~~~~~~~~~~~~            \rightarrow 2s^22p^53d \ ^3P^o_2 $&2 &4460.30 &27193.78   &26689.49  &-10085.80    \\

$2s^22p^53d\ ^3P^o_1           \rightarrow 2s^22p^53d \ ^3P^o_2 $&2 &6849.31 &15984.06   &15711.27  &-5455.80    \\

\hline \\
Co XVIII \\
$2s^22p^6  \ ^1S_0         \rightarrow 2s^22p^53s \ ^3P^o_1$            &1 &15.43&6481449.61  &6483622.64 &-43460.60    \\
$~~~~~~~~~~~~~~~           \rightarrow 2s^22p^53s \ ^1P^o_1$            &1 &15.16&6603472.72  &6599388.48 &81684.80    \\   
$~~~~~~~~~~~~~~~           \rightarrow 2s^22p^53d \ ^3P^o_1$            &1 &14.04&7130003.64  &7130057.70 &-1081.20    \\
$~~~~~~~~~~~~~~~           \rightarrow 2s^22p^53d \ ^3D^o_1$            &1 &13.86&7220232.64  &7219522.98 &14193.20    \\
$~~~~~~~~~~~~~~~           \rightarrow 2s^22p^53d \ ^1P^o_1$            &1 &13.63&7346134.12  &7341253.33 &97615.80    \\

$2s^22p^53s \ ^3P^o_1      \rightarrow 2s^22p^53s \ ^1P^o_1$            &1 &873.36&122023.11   &115765.84  &-125145.40    \\   
$~~~~~~~~~~~~~~~~~~        \rightarrow 2s^22p^53d \ ^3P^o_1$            &1 &155.25&648554.03   &646435.06  &-42379.40    \\
$~~~~~~~~~~~~~~~~~~        \rightarrow 2s^22p^53d \ ^3D^o_1$            &1 &136.44&738783.03   &735900.34  &-57653.80    \\
$~~~~~~~~~~~~~~~~~~        \rightarrow 2s^22p^53d \ ^1P^o_1$            &1 &116.72&864684.51   &857630.69  &-141076.40    \\

$2s^22p^53s \ ^1P^o_1      \rightarrow 2s^22p^53d \ ^3P^o_1$            &1 &188.82&526530.92   &530669.22  &82766.00    \\
$~~~~~~~~~~~~~~~~~~        \rightarrow 2s^22p^53d \ ^3D^o_1$            &1 &161.70&616759.92   &620134.50  &67491.60    \\
$~~~~~~~~~~~~~~~~~~        \rightarrow 2s^22p^53d \ ^1P^o_1$            &1 &134.73&742661.40   &741864.85  &-15931.00    \\

$2s^22p^53d \ ^3P^o_1      \rightarrow 2s^22p^53d \ ^3D^o_1$            &1 &1126.12&90229.00    &89465.28   &-15274.40    \\
$~~~~~~~~~~~~~~~~~~        \rightarrow 2s^22p^53d \ ^1P^o_1$            &1 &470.36 &216130.48   &211195.63  &-98697.00    \\
                                                                       
$2s^22p^53d \ ^3D^o_1      \rightarrow 2s^22p^53d \ ^1P^o_1$            &1 &807.75&125901.48   &121730.35  &-83422.60    \\
\end{tabular}   
\end{ruledtabular}
\label{tab2}
\end{table*}

In Table \ref{tab3}, we report the results of the $\alpha$ sensitivity coefficients for the Cr XV and Ni XIX 
ions. For both the ions, we have considered 25 possible transitions to estimate the $q$ parameters. In case
of Cr XV, the transition $2s^22p^53d \ ^3P^o_0 \rightarrow 2s^22p^53d \ ^3P^o_2$ lie in the optical region
whereas all other transitions fall in the UV range. Out of 25 considered transitions in Cr XV, $8$ of them
have positive $q$-values and the rest have negative $q$-values as given in Table \ref{tab3}. In this ion,
the largest positive and negative $q$-parameters correspond to the transitions 
$2s^22p^53s \ ^3P^o_1 \rightarrow 2s^22p^53d \ ^3P^o_0$ and $2s^22p^53s \ ^3P^o_2 \rightarrow 2s^22p^53d \ ^3P^o_0$ 
respectively. At last, we also present the results for the Ni XIX ion in Table \ref{tab3}. Similar to the case of 
Cr XV, Ni XIX ion also has one optical transition $2s^22p^53s \ ^3P^o_2 \rightarrow 2s^22p^53s \ ^3P^o_1$ and 
rest of the transitions fall in the UV region. It shows that the $q$-values in this ion have the similar trends
like in Cr XV. We have $9$ positive and $16$ negative $q$-coefficients with the largest positive and 
negative $q$-values as $104044.80 \ cm^{-1}$ and $-149579.20 \ cm^{-1}$ respectively. 

Analogous to the Fe XVII ion, the possible anchor lines in the case of Cr XV correspond to 
wavelengths $19.01$ \AA, 7572.31 \AA~ and 11079.10 \AA ~with the $q$-values $-1342.60$, $-956.40$ and $920.60$, 
respectively, in $cm^{-1}$.
Moreover in the Ni XIX ion, the transitions  $2s^22p^6 \ ^1S_0 \rightarrow 2s^22p^53d \ ^3P^o_2$, 
$2s^22p^53s \ ^3P^o_2 \rightarrow 2s^22p^53s \ ^1P^o_1$ and $2s^22p^53s \ ^3P^o_0 \rightarrow 2s^22p^53s \ ^3P^o_1$ which 
lie in the EUV, optical and NIR region respectively, have small $q$-values and can be chosen as the anchor lines.

\begin{table*}[t]
\caption{Sensitivity $q$ coefficients (in $cm^{-1}$) for the Cr XV and Ni XIX ions using the CCSD method.
The frequencies $\omega (+0.025)$ and $\omega (+0.025)$ are given as absolute values.}
\begin{ruledtabular}
\begin{tabular}{lccccc}
\multicolumn{1}{c}{Transitions}&
\multicolumn{1}{c}{$J_f$}&
\multicolumn{1}{c}{$\lambda (\AA)$}&
\multicolumn{1}{c}{$\omega(+0.025)$}&
\multicolumn{1}{c}{$\omega(-0.025)$}&
\multicolumn{1}{c}{$q$}\\
\hline\\ 
Cr XV \\
$2s^22p^6 \ ^1S_0               \rightarrow 2s^22p^53s \ ^3P^o_2 $&2 &21.21&4710372.93  &4711725.19 &-27045.20     \\   
$~~~~~~~~                       \rightarrow 2s^22p^53s \ ^1P^o_1 $&1 &21.15&4727989.51  &4729293.95 &-26088.80     \\
$~~~~~~~~                       \rightarrow 2s^22p^53s \ ^3P^o_1 $&1 &20.86&4798109.43  &4795838.96 &45409.40     \\   
$~~~~~~~~                       \rightarrow 2s^22p^53d \ ^3P^o_1 $&1 &19.01&5261218.88  &5261286.01 &-1342.60     \\
$~~~~~~~~                       \rightarrow 2s^22p^53d \ ^3P^o_2 $&2 &18.97&5273946.73  &5273762.90 &3676.60     \\

$2s^22p^53s \ ^3P^o_2           \rightarrow 2s^22p^53s \ ^1P^o_1 $&1 &7572.31&17616.58    &17568.76   &-956.40     \\
$~~~~~~~~                       \rightarrow 2s^22p^53s \ ^3P^o_0 $&0 &1431.02&74775.43    &71106.67   &-73375.20     \\
$~~~~~~~~                       \rightarrow 2s^22p^53s \ ^3P^o_1 $&1 &1267.33&87736.50    &84113.77   &-72454.60     \\   
$~~~~~~~~                       \rightarrow 2s^22p^53d \ ^3P^o_0 $&0 &185.47 &541564.39   &540494.06  &-21406.60     \\
$~~~~~~~~                       \rightarrow 2s^22p^53d \ ^3P^o_2 $&1 &183.44 &550845.95   &549560.82  &-25702.60     \\
$~~~~~~~~                       \rightarrow 2s^22p^53d \ ^3P^o_2 $&2 &179.64 &563573.80   &562037.71  &-30721.80     \\

$2s^22p^53s \ ^1P^o_1           \rightarrow 2s^22p^53s \ ^3P^o_0 $&0 &1764.48&57158.85    &53537.91   &-72418.80     \\
$~~~~~~~~                       \rightarrow 2s^22p^53s \ ^3P^o_1 $&1 &1522.07&70119.92    &66545.01   &-71498.20     \\   
$~~~~~~~~                       \rightarrow 2s^22p^53d \ ^3P^o_0 $&0 &190.13 &523947.81   &522925.30  &-20450.20     \\
$~~~~~~~~                       \rightarrow 2s^22p^53d \ ^3P^o_1 $&1 &187.99 &533229.37   &531992.06  &-24746.20     \\
$~~~~~~~~                       \rightarrow 2s^22p^53d \ ^3P^o_2 $&2 &184.01 &545957.22   &544468.95  &-29765.40     \\
                                                                                                       
$2s^22p^53s \ ^3P^o_0           \rightarrow 2s^22p^53s \ ^3P^o_1 $&1 &11079.10&12961.07    &13007.10   &920.60       \\   
$~~~~~~~~                       \rightarrow 2s^22p^53d \ ^3P^o_1 $&1 &210.42&476070.52   &478454.15  &47672.60     \\
$~~~~~~~~                       \rightarrow 2s^22p^53d \ ^3P^o_2 $&2 &205.43&488798.37   &490931.04  &42653.40     \\

$2s^22p^53s \ ^3P^o_1           \rightarrow 2s^22p^53d \ ^3P^o_0 $&0 &217.27&453827.89   &456380.29  &51048.00     \\
$~~~~~~~~                       \rightarrow 2s^22p^53d \ ^3P^o_1 $&1 &214.49&463109.45   &465447.05  &46752.0     \\
$~~~~~~~~                       \rightarrow 2s^22p^53d \ ^3P^o_2 $&2 &209.31&475837.30   &477923.94  &41732.80     \\

$2s^22p^53d \ ^3P^o_0           \rightarrow 2s^22p^53d \ ^3P^o_1 $&1 &16747.61&9281.56     &9066.76    &-4296.00     \\
$~~~~~~~~                       \rightarrow 2s^22p^53d \ ^3P^o_2 $&2 &5715.26 &22009.41    &21543.65   &-9315.20     \\

$2s^22p^53d \ ^3P^o_1           \rightarrow 2s^22p^53d \ ^3P^o_2 $&2 &8676.03 &12727.85    &12476.89   &-5019.20     \\
\hline \\ 
Ni XIX \\
$2s^22p^6\ ^1S_0                \rightarrow 2s^22p^53s \ ^3P^o_2 $&2 &14.07&7100997.25  &7103565.26 &-51360.20     \\   
$~~~~~~~~                       \rightarrow 2s^22p^53s \ ^1P^o_1 $&1 &14.03&7121845.61  &7124365.04 &-50388.60     \\
$~~~~~~~~                       \rightarrow 2s^22p^53s \ ^3P^o_1 $&1 &13.78&7265582.55  &7260715.41 &97342.80     \\   
$~~~~~~~~                       \rightarrow 2s^22p^53d \ ^3P^o_1 $&2 &12.74&7830543.23  &7830327.75 &4309.60     \\
$~~~~~~~~                       \rightarrow 2s^22p^53d \ ^3P^o_2 $&1 &12.80&7811084.03  &7811160.03 &-1520.00     \\

$2s^22p^53s \ ^3P^o_2           \rightarrow 2s^22p^53s \ ^1P^o_1 $&1 &5767.01&20848.36   &20799.78  &-971.60     \\
$~~~~~~~~                       \rightarrow 2s^22p^53s \ ^3P^o_0 $&0 &702.05 &150566.74  &143087.78 &-149579.20     \\
$~~~~~~~~                       \rightarrow 2s^22p^53s \ ^3P^o_1 $&1 &654.27&164585.30  &157150.15 &-148703.0     \\   
$~~~~~~~~                       \rightarrow 2s^22p^53d \ ^3P^o_0 $&0 &144.36&696739.51  &694506.60 &-44658.20     \\
$~~~~~~~~                       \rightarrow 2s^22p^53d \ ^3P^o_2 $&2 &134.80&729545.98  &726762.49 &-55669.8     \\
$~~~~~~~~                       \rightarrow 2s^22p^53d \ ^3P^o_2 $&1 &142.36&710086.78  &707594.77 &-49840.20     \\

$2s^22p^53s \ ^1P^o_1           \rightarrow 2s^22p^53s \ ^3P^o_0 $&0 &799.36&129718.38  &122288.00 &-148607.60     \\
$~~~~~~~~                       \rightarrow 2s^22p^53s \ ^3P^o_1 $&1 &738.00&143736.94  &136350.37 &-147731.40     \\   
$~~~~~~~~                       \rightarrow 2s^22p^53d \ ^3P^o_0 $&0 &148.07&675891.15  &673706.82 &-43686.60     \\
$~~~~~~~~                       \rightarrow 2s^22p^53d \ ^3P^o_1 $&2 &138.02&708697.62  &705962.71 &-54698.20     \\
$~~~~~~~~                       \rightarrow 2s^22p^53d \ ^3P^o_2 $&1 &145.96&689238.42  &686794.99 &-48868.60     \\

$2s^22p^53s \ ^3P^o_0           \rightarrow 2s^22p^53s \ ^3P^o_1 $&1 &9615.38&14018.56   &14062.37  &876.20       \\   
$~~~~~~~~                       \rightarrow 2s^22p^53d \ ^3P^o_1 $&2 &166.83&578979.24  &583674.71 &93909.40     \\
$~~~~~~~~                       \rightarrow 2s^22p^53d \ ^3P^o_2 $&1 &178.57&559520.04  &564506.99 &99739.00     \\

$2s^22p^53s \ ^3P^o_1           \rightarrow 2s^22p^53d \ ^3P^o_0 $&0 &185.23&532154.21  &537356.45 &104044.80     \\
$~~~~~~~~                       \rightarrow 2s^22p^53d \ ^3P^o_1 $&2 &169.78&564960.68  &569612.34 &93033.20     \\
$~~~~~~~~                       \rightarrow 2s^22p^53d \ ^3P^o_2 $&1 &181.95&545501.48  &550444.62 &98862.80     \\

$2s^22p^53d \ ^3P^o_0           \rightarrow 2s^22p^53d \ ^3P^o_1 $&2 &2035.21&32806.47   &32255.89  &-11011.60     \\
$~~~~~~~~                       \rightarrow 2s^22p^53d \ ^3P^o_2 $&1 &10272.21&13347.27   &13088.17  &-5182.00     \\

$2s^22p^53d \ ^3P^o_1           \rightarrow 2s^22p^53d \ ^3P^o_2 $&1 &2538.07&19459.20   &19167.72  &5829.60     \\

\end{tabular}   
\end{ruledtabular}
\label{tab3}
\end{table*}

\section{Conclusion}

We have implemented an equation-of-motion coupled-cluster method to calculate
the excited states of a closed-shell atomic system in the four-component relativistic
framework that preserves spherical symmetric properties explicitly. The method has been
employed to calculate the excitation energies of four different highly charged ions 
that are of astrophysical interest. Our calculations are 
very accurate as compared to their corresponding experimental values. 
The development will be very useful to study a variety of atomic properties of many 
atomic systems for which high-precision calculations are in demand. The present
method can adequately address the role of the relativistic and the electron correlation effects
scrupulously to achieve high precision results to explain many physical problems 
of modern research interest. To illustrate its potential application, we employ
the above method to calculate excitation energies for different values of the
fine structure constant in a number of transitions in the considered ions and 
estimate the relativistic sensitivity coefficients that are of vested interest
in the investigation of temporal variation of the fine structure constant using the
laboratory astrophysics method. We found large sensitivity coefficients of opposite
signs that may be of very crucial information to be analyzed for the detection of 
enhanced drifts in the course of searching possible variation in the fine structure constant.

\section*{Acknowledgment}

We are grateful to B. P. Das and D. Mukherjee for many useful discussions on the
theory during the implementation of the code. We thank M. Kallay for providing us
the diagonalization program used in the code to obtain lower roots of a non-symmetric
matrix. The calculations were carried out using the PRL 3TFLOP HPC cluster, Ahmadabad.

\end{document}